\documentclass{article}

\usepackage{arxiv}

\usepackage{multirow}
\usepackage{array}
\usepackage[ruled,vlined]{algorithm2e}

\usepackage{xcolor}

\usepackage[utf8]{inputenc} 
\usepackage[T1]{fontenc}    
\usepackage{hyperref} 
\usepackage{scalerel,stackengine}
\stackMath
\newcommand\reallywidehat[1]{%
\savestack{\tmpbox}{\stretchto{%
  \scaleto{%
    \scalerel*[\widthof{\ensuremath{#1}}]{\kern-.6pt\bigwedge\kern-.6pt}%
    {\rule[-\textheight/2]{1ex}{\textheight}}
  }{\textheight}%
}{0.5ex}}%
\stackon[1pt]{#1}{\tmpbox}%
}

\usepackage{graphicx}
\usepackage[export]{adjustbox}
\usepackage{xcolor}
\usepackage{subfigure}
\usepackage{colortbl}
\usepackage{paralist}
\usepackage{bbm}
\usepackage{url}            
\usepackage{booktabs}       
\usepackage{amsfonts}       
\usepackage{nicefrac}       
\usepackage{microtype}      
\usepackage{lipsum}	
\usepackage{xcolor}
\title{Fringe News Networks: Dynamics of US News Viewership following the 2020 Presidential Election}

\author{
Ashiqur R. KhudaBukhsh\thanks{Ashiqur R. KhudaBukhsh and Rupak Sarkar  are equal contribution first authors.} \\
  \small{Carnegie Mellon University} \\
  \texttt{akhudabu@cs.cmu.edu} \\
  \And
  Rupak Sarkar$^*$\\
  \small{Maulana Abul Kalam Azad University of Technology}\\
  \texttt{rupaksarkar.cs@gmail.com } \\
\And
Mark S. Kamlet \\
  \small{Carnegie Mellon University}\\
  \texttt{kamlet@cmu.edu} \\
 \And
Tom M. Mitchell \\
  \small{Carnegie Mellon University}\\
  \texttt{tom.mitchell@cs.cmu.edu} \\
}

\renewcommand{\headeright}{}

\begin{document}
\maketitle

\begin{abstract}

The growing political polarization of the American electorate over the last several decades has been widely studied and documented. During the administration of President Donald Trump, charges of “fake news” made social and news media not only the means but, to an unprecedented extent, the topic of political communication.  Using data from before the November 3rd, 2020 US Presidential election, recent work has demonstrated the viability of using YouTube’s social media ecosystem to obtain insights into the extent of US political polarization as well as the relationship between this polarization and the nature of the content and commentary provided by different US news networks.  With that work as background, this paper looks at the sharp transformation of the relationship between news consumers and here-to-fore ``fringe'' news media channels  in the 64 days between the US presidential election and the violence that took place at US Capitol on January 6$^\mathit{th}$.   This paper makes two distinct types of contributions. The first is to introduce a novel methodology to analyze large social media data to study the dynamics of social political news networks and their viewers.  The second  is to provide insights into what actually happened regarding US political social media channels and their viewerships during this volatile 64 day period.

\end{abstract}


\section{Introduction}
The growing political polarization of the American electorate over the last several decades has been widely studied and documented~\cite{poole1984polarization, layman2010party,mccarty2016polarized, baldwin2016past,mcconnell2017research, demszky-etal-2019-analyzing, darwish2019quantifying, khudabukhsh2020dont}.  
During the administration of President Donald Trump, charges of ``fake news'' made social and news media not only the means but, to an unprecedented extent, the topic of political communication ~\cite{ross2018discursive}. At the same time, the partisan and ideological divergences across viewers of major US television networks increasingly became mirrored by the content and commentary which these networks provided~\cite{NYTimeEvilTwin, bozell2004weapons, gil2012selective, hyun2016agenda}.

Recent study has demonstrated the viability of using YouTube's social media ecosystem to obtain insights on both of these considerations~\cite{khudabukhsh2020dont}.  
That study focused on viewer responses to videos from the ``big three'' cable news networks--namely CNN, Fox News, and MSNBC. While that study did give cursory attention to One America News Network (OANN), the focus on CNN, Fox News, and MSNBC was logical given the size of these news networks' YouTube viewership at the time. The relative size of these networks' YouTube viewing audiences remained approximately stable through November 3$^\mathit{rd}$, 2020 when the US Presidential election was held. But, as is described shortly, what happened next was anything but stable.  

This paper makes two distinct types of contributions. The first is to introduce a novel methodology to analyze large social media data to study the dynamics of news networks and their viewers. This methodology applies a variety of state-of-the-art methods including machine translation; specialized deep network language models trained on different portions of social media text; and cloze tests on those distinct language models to study the difference in opinions across these different subcommunities. Taken together, these methods provide a complementary and corroborative portrayal of the dynamics of viewers' expressed social media opinions; the nature of the content and commentary set forth by the different news networks; and the changes in the alignment between the two. That the results reported in this paper could be obtained so quickly after the event highlights this methodology's potential power and usefulness, and suggest opportunities that may well also be applicable for analysis of other such time-series data.

The second type of contribution is to provide insights into what actually happened regarding US political social media channels and their viewerships during the 64 days between the US Presidential election on November 3$^\mathit{rd}$, 2020 and the entry into the US Capitol of violent demonstrators on January 6$^\mathit{th}$, 2021. To briefly provide necessary background, a precipitating event to what subsequently took place occurred during the evening of November 4$^\mathit{th}$, as election results were beginning to come in. Following an unexpectedly strong showing for the President in Florida compared with virtually all prior polling, the mood in the White House was upbeat\footnote{\url{https://www.nytimes.com/2020/11/04/us/politics/trump-fox-news-arizona.html}}.
But, as reported by the New York Times ``$\ldots$[the] mirage of victory was pierced when Fox News called Arizona for former Vice President Joseph R. Biden Jr. at 11:20 p.m., with just 73 percent of the state’s vote counted.'' 
Fox News made this call before other national networks had done so. In fact, it wasn't until November 12$^\mathit{th}$, nine days after Election Day, that the other networks with decision desks — NBC, ABC, CBS and CNN — called the state for Biden too.

As reported by The Times, ``Mr. Trump and his advisers erupted at the news. If it was true that Arizona was lost, it would call into doubt on any claim of victory the president might be able to make.'' 
\footnote{In response to White House criticism, Fox interviewed at approximately 1 a.m. on November 4$^\mathit{th}$ Arnold Mishkin, director of the Fox News Decision Desk. Mr. Mishkin stated: ``We’re four standard deviations from being wrong. And, I’m sorry, we’re not wrong in this particular case.''}
What ensued for Mr. Trump, again according to the Times, "was a night of angry calls to Republican governors$\ldots$leading to a middle-of-the-night presidential briefing in which he made a reckless and unsubstantiated string of remarks about the democratic process. Standing in the East Room at 2:30 a.m., he dismissed the election as a `fraud'." 

With this as background, this paper's analyses find the following key conclusions. 
\begin{itemize}
\item Following November 3$^\mathit{rd}$, there was a notable loss of Fox's YouTube channel's market share, and the departure of previously loyal Fox viewers to what here-to-fore were considered fringe networks (OANN, Newsmax, and Blaze). As an example, the viewership of Newsmax increased by over a factor of seven from the pre-election period to January 6$^\mathit{th}$, 2021.

\item Compared to networks such as CNN, MSNBC, and FoxNews, we find that the networks OANN, Newsmax, and Blaze had more features of being ``echo chambers,'' in the sense that their viewerships more nearly uniformly agreed with what they were watching, with lower proportions of their viewerships critiquing what was being presented to them. 

\item We find that viewer opinion about the legitimacy of the election is polarized into two groups, with viewers of MSNBC, CNN, and Fox News far more in agreement that Biden should be considered ``president-elect'' than OANN, Newsmax, and Blaze. In a similar vein, OANN and Newsmax are strong outliers in terms of usage of the trigram ``stop the steal.''

\item Based on cloze tests~\cite{taylor1953cloze} using the probe \emph{The biggest problem of American is} \texttt{[MASK]}, training a language model~\cite{devlin2018bert} based on the comments provided by MSNBC viewers, the top three answers are ``Trump'', ``COVID,'' and ``unemployment'', while a language model trained on comments provided by OANN yields ``communism,'' ``corruption,'' and ``socialism.'' The other networks fall into positions along this continuum that are consistent with expectations. A similar behavior is observed when cloze tests are employed to analyze who won the election. These findings are corroborated further with a Natural Language Inference algorithm.   


\item Using a machine translation based method presented in~\cite{khudabukhsh2020dont} that quantifies the differences between large-scale social media discussion corpora, each channel's viewership is assigned its own \emph{language} and the similarities between the languages of any two channels can be quantified. The language of the viewership of Trump's own individual YouTube channel is most similar to the language of the viewership of Newsmax, followed by OANN, followed by Blaze, followed by Fox, followed by CNN, followed by MSNBC. 
\end{itemize}

\section{Data Set}

Our data set considers official YouTube channels of six US cable news networks listed in Table~\ref{tab:network}, and consists of: subscription counts of these YouTube channels; comments posted by viewers of individual news videos posted by each channel;  "likes" and "dislikes" associated with each of with these videos; and news video transcripts\footnote{Of these six news networks, we refer to Blaze TV, OANN, and Newsmax as fringe news networks due to their relatively homogeneous audience and limited reach compared to the big three.}. In addition to these six YouTube channels, we consider the official YouTube channel of the 45$^\mathit{th}$ US President, Donald J. Trump. We used the publicly available YouTube API to download comments, and video ``likes'' and ``dislikes'' information. Apart from CNN, for each news video, we also extracted video transcripts using a Python package\footnote{\url{https://pypi.org/project/youtube-transcript-api/}}. The package did not give reliable results for CNN, hence we omit CNN in our analyses on the news transcripts (presented in Section~\ref{sec:trigram}).  

Our analyses primarily focus on two  non-overlapping time intervals. We denote the time interval of 31$^\mathit{st}$ August, 2020 to 2$^\mathit{nd}$ November, 2020, i.e., the 64 days leading up to the 2020 US election, as $\mathcal{T}_\mathit{before}$. $\mathcal{T}_\mathit{after}$ refers to the time interval starting from November 3$\mathit{rd}$, 2020 to January 5$\mathit{th}$, 2021.      Starting from 31$^\mathit{st}$ August, 2020 to 5$^\mathit{th}$ January, 2021. We denote the combined time interval of these 128 days as $\mathcal{T}_\mathit{128}$. Our data set consists of 14,557,966 comments on 11,964 videos posted by 2,278,034 users.


\begin{table}[htb]
{

\begin{center}
     
\begin{tabular}{|l | c  |c | c |}
    \hline
    YouTube Channel & \#Subscribers  &\#Videos during $\mathcal{T}_\mathit{128}$ & Total \#Comments \\
    \hline                                 
   CNN    &  11.7M  &824 &3,368,178
 \\
    \hline
   Fox News  & 6.71M  &2,066 &4,059,446 \\
    \hline
  MSNBC  &  3.97M  & 3,890 &2,776,968 \\
    \hline
OANN   &  1.36M  & 1,728 &427,908 \\
  \hline
Newsmax   & 1.77M  &746 &971,617 \\
  \hline 
Blaze TV   & 1.34M  &518 &634,650 \\
  \hline
  Donald J. Trump   & 2.68M  & 2,212 &2,382,821 \\
  \hline
    \end{tabular}

\vspace{0.2cm}
\caption{List of news networks considered. Video counts during $\mathcal{T}_\mathit{128}$ reflect the number of videos uploaded  on or before 5$^\mathit{th}$ January 2021 starting from 31$^\mathit{st}$ August 2020.}
\label{tab:network}
\end{center}
}

\end{table}

\section{Related Work}

Previous research on US cable news reported divergent views both in audience and in content~\cite{gil2012selective, hyun2016agenda}. However, these works primarily relied on surveys and were restricted to the television medium without considering these channels' YouTube presence and therefore were unable to tap into user comments and interactions. In terms of the nature of our data set, our work is closest to~\cite{khudabukhsh2020dont} in its use of comments on YouTube news videos of major US cable news networks. We also leverage the linguistic framework and a measure to estimate viewership agreement from this work. Our work contrasts with~\cite{khudabukhsh2020dont} in the following key ways: (1) our focus on an important (and timely) and non-overlapping period of 64 days prior and after the 2020 US election; (2) our emphasis on three fringe news networks, two of which were (Newsmax and Blaze TV) previously ignored in~\cite{khudabukhsh2020dont} and one briefly analyzed; and (3) our use of a wider variety of NLP tools in analyzing a broader range of research questions rather than presenting a quantifiable framework to gauge linguistic polarization. 

Previous work on deplatforming has analyzed effects of large-scale bans of communities on other social media platforms such as Reddit~\cite{chandrasekharan2017you}. Our work on analyzing the migration of Fox News viewers to Newsmax adds a subtle nuance that in this case users are not being deplatformed by the platform owners. It is rather (potentially) triggered by calling the election as per the Associated Press. Echo chambers in social media is a widely studied topic~\cite{jacobson2016open, boutyline2017social,gillani2018me}. Our work is similar to past work on analyzing the presence of echo chambers~\cite{lima2018inside} in conservative forums with a key distinction that our choice of platform is heavily mainstream.   

Our work draws inspiration from several recent NLP contributions analyzing political corpora~\cite{electionKhudaBukhsh,khudabukhsh2020dont} or misinformation~\cite{hossain-etal-2020-covidlies}. For instance,~\cite{electionKhudaBukhsh} presented an application of language models~\cite{devlin2018bert} to mine insights and aggregate opinions using language models fine-tuned on an Indian political social media data set. Similarly,~\cite{hossain-etal-2020-covidlies} presents a link between stance detection and the entailment literature in the context of detecting COVID-19 misinformation.   Instead of methodologically advancing these techniques, in this work we demonstrate the synergy between these methods on a critical domain of political crisis. 

\section{Results}

We present a road map of our results section with our research questions and relevant sections.  

\subsection{Research questions}
\emph{\textbf{RQ 1}: Did different networks pursue different approaches toward accepting and presenting the election outcomes e.g., if there was widespread voter fraud? } (discussed in Section~\ref{sec:trigram})\\
\noindent\emph{\textbf{RQ 2}: Did the audience of different networks exhibit different attitudes toward accepting and presenting the election outcomes?} (discussed in Section~\ref{sec:trigram},~\ref{sec:cloze})\\
\emph{\textbf{RQ 3}: Were there any shifts in viewership engagement of news networks post election?} (discussed in Section~\ref{sec:engagement})\\
\noindent\emph{\textbf{RQ 4}: Was there any systematic migrations from mainstream media outlets to fringe media outlets?}~(discussed in Section~\ref{sec:fox}) \\
\noindent\emph{\textbf{RQ 5}: Based on the comments on the viewed videos, which news networks were ``linguistically most similar'' to those of President Trump's YouTube channel?} (discussed in Section~\ref{sec:translation})\\

\subsection{The story of two trigrams}\label{sec:trigram}

\emph{\textbf{RQ 1}: Did different networks pursue different approaches toward accepting and presenting the election outcomes e.g., if there was widespread voter fraud?}\\
\noindent\emph{\textbf{RQ 2}: Did the audience of different networks exhibit different attitudes toward accepting and presenting the election outcomes?}

We start with a simple analysis involving two short phrases to characterize (1) the portrayal of the election outcome across different news networks, and (2) how the viewership of the said networks responded during this period. 

Our selected phrases are ``President-elect Biden'' and ``stop the steal'' (and a few high-frequency variants of these -- e.g., ``President-elect [$\mathit{wildcard}$] Biden'' to make room for Joseph or Joe or Joseph R.). We examine the first phrase using the video transcripts.  We argue that after November 7$^\mathit{th}$ 2020, when the Associated Press called the election for Biden, any reference to President-elect Biden in any news video indicates support for the legitimacy of the Biden victory\footnote{Of course, there could be counter-examples  -- for instance, an anchor saying ``I am never going to refer to him as President-elect Biden until Supreme Court hears the case''. We manually inspected 100 randomly sampled unique references across 100 videos and confirm that is not the case.}. We examine the usage of our second phrase on our data set consisting of user comments on news videos. The choice of our next phrase is guided by ``stop the steal'' protests aimed at discrediting the 2020 election outcome\footnote{In fact, this particular phrase has a history that goes beyond the 2020 election; Trump advisor Roger Stone ran an organization~\cite{gitlinvoter} with a name identical to this phrase to detect voter fraud in the 2016 US election.}. In this case, our intuition is if a user comment mentions this phrase (or some variant of it), it is highly likely that the user is expressing a belief that the election is fraudulent\footnote{\url{https://www.nbcnews.com/tech/tech-news/facebook-bans-all-stop-steal-content-n1253809}.}.

Through the usage pattern of our first phrase (``President-elect Biden''), we now estimate the overall stance of a news network across the individual videos hosted in its official YouTube channel. Let the indicator function $\mathbbm{I} (v,$ ``President-elect~Biden"$)$ returns 1 if the said phrase (or some variant of it) is mentioned at least once in the video transcript of $v$ and returns 0 otherwise. Similarly, let the indicator function $\mathbbm{I} (v,$ ``Biden"$)$ returns 1 if ``Biden'' is mentioned at least once in the video transcript of $v$ and returns 0 otherwise. For a given channel and videos posted within November 7$^{th}$ 2020 and January 5$^{th}$ 2021, we compute the following factor:\\ \large{$\frac{\Sigma_i \mathbbm{I}(v^i, ``President-elect~Biden") }{\Sigma_i \mathbbm{I}(v^i, ``Biden")}$}.
\normalsize
Table~\ref{tab:preselect}
lists the value of our measure across each news network. We note that, while the two mainstream media outlets exhibit comparable mentions of the phrase ``President-elect'', the three conservative fringe networks show remarkably fewer mentions of this term indicating a possible stance of not accepting the official outcome of the election. Of the three big networks, Fox News is a well-known conservative network. This measure further indicates the possibility that a fringe network may afford to present a narrower view of an event than its mainstream media outlets catering to a wider audience and yet enjoy substantial audience approval and engagement (audience approval and engagement results are presented in Section~\ref{sec:engagement}).

\begin{table}[htb]
{
\begin{center}

\begin{tabular}{|l | c |}
    \hline
    YouTube Channel & Measure \\
    \hline                                 
   CNN   & - \\
    \hline
   Fox News & \textbf{29.1\%} \\
    \hline
  MSNBC  & 28.8\% \\
    \hline
  OANN  & 2.7\% \\
  \hline
  Newsmax  & 6.5\% \\
  \hline 
  Blaze TV  & 13.9\% \\
    \hline
  
    \end{tabular}

\end{center}
}
\caption{Analysis of the overall stance toward accepting the election outcome of Biden being the President-elect across different news networks. Percentages shown are the percentage of times that a news video mentioning Biden refers to him as "President elect."  
These results indicate that both mainstream media outlets Fox News and MSNBC  referred to Biden as President-elect relatively more than the fringe media outlets.}
\label{tab:preselect}
\end{table}

We now answer our second research question using the ``stop the steal" trigram. Table~\ref{tab:stopthesteal} presents the frequency-based rank of the trigram over the discussion data set of each of the news networks. In order to ensure that these rankings are comparable across news networks, each of the corpora has identical number of tokens. A relatively higher rank of this phrase in network $\mathit{network}_i$ over network $\mathit{network}_j$ indicates that the phrase is relatively more popular in $\mathit{network}_i$. Table~\ref{tab:stopthesteal} shows that while this trigram is considerably popular across all six channels, OANN and Newsmax particularly stand out.  

\begin{table}[htb]
{

\begin{center}

\begin{tabular}{|l | c  |}
    \hline
    YouTube channel &  Rank  \\
    \hline                                 
   CNN     & 111 \\
    \hline
   Fox News  & 134\\
    \hline
  MSNBC   & 123 \\
    \hline
  OANN   & 75 \\
  \hline
  Newsmax   & 63 \\
  \hline 
  Blaze TV   & 111 \\
    \hline
  
    \end{tabular}
\vspace{0.2 cm}
\caption{Analysis of the ``stop the steal'' phrase in comments on news videos across news networks. Table~\ref{tab:stopthesteal} presents the frequency-based rank of the trigram over the discussion data set of each of the news networks. }
\label{tab:stopthesteal}
\end{center}
}
\end{table}

\subsection{Post Election Engagement Shift}\label{sec:engagement}

\emph{\textbf{RQ 3}: Were there any shifts in viewership engagement of news networks post election?}\\

We investigate this research question through three signals: (1) video likes and dislikes; (2) average comment count; and (3) news network subscriber count.  

\subsubsection{Video likes and dislikes}

Following \cite{khudabukhsh2020dont}, we use the same viewership disagreement measure to estimate disagreement in a network. Let $v_\mathit{like}$ and $v_\mathit{dislike}$ denote the total number of likes and dislikes received by a given video $v$. Let for a given channel $\mathcal{C}$, $\mathbbm{I}(v^i, \mathcal{C}, \mathcal{T})$ returns 1 if video $v^i$ is uploaded to $\mathcal{C}$ within duration $\mathcal{T}$, otherwise it returns 0. The disagreement factor of a channel $\mathcal{C}$ for a given time duration $\mathcal{T}$ is thus calculated as \large{$\frac{\Sigma_i \mathbbm{I}(v^i, \mathcal{C}, \mathcal{T}) \frac{v^i_\mathit{dislike}}{v^i_\mathit{dislike} + v^i_\mathit{like}}}{\Sigma_i \mathbbm{I}(v^i, \mathcal{C}, \mathcal{T})}$}.
\normalsize
The interpretation of a low value of this measure is overall, videos are generally liked by substantially more viewers than disliked in the channel. A higher value indicates mixed user response with an increasing fraction of disapproving viewership. As a nice property of this measure, ~\cite{khudabukhsh2020dont} further points that the ratio $\frac{v_{\mathit{dislike}}}{v_{\mathit{dislike}} + v_{\mathit{like}}}$ for an individual video and the overall measure are both bounded within [0,1] and one arbitrarily heavily liked or disliked video can at most influence the overall average by $\frac{1}{n}$ where $n$ is the total number of videos uploaded in that particular duration (as shown in Table~\ref{tab:agreement}, the minimum value for $n$ in our case is 294).

Table~\ref{tab:agreement} presents the disagreement factor for each channel for time duration $\mathcal{T}_\mathit{before}$ and $\mathcal{T}_\mathit{after}$ and the difference in disagreement (denoted by $\Delta_{\mathit{disagreement}}$) obtained by subtracting the disagreement in $\mathcal{T}_\mathit{after}$ from  $\mathcal{T}_\mathit{before}$. A positive $\Delta_{\mathit{disagreement}}$ indicates that the channel has gained popularity while a negative value indicates a decline in   popularity. We note that apart from Fox News, $\Delta_{\mathit{disagreement}}$ is within $\pm 0.03$ for all other news networks.     

\begin{table}[htb]
{

\begin{center}

\begin{tabular}{|l | c | c | c |}
    \hline
    YouTube channel & $\mathcal{T}_\mathit{before}$ & $\mathcal{T}_\mathit{after}$ & $\Delta_\mathit{disagreement}$\\
    \hline                                 
   CNN   &  0.20	      &  0.17 & +0.03 \\
    \hline
   Fox News &  0.18  & \textbf{0.28} & \textbf{-0.10} \\
    \hline
  MSNBC  &  0.10  & 0.09 & +0.01  \\
    \hline
  OANN  &  0.02    & 0.02 & 0 \\
  \hline
  Newsmax  &  0.01 & 0.02 & -0.01  \\
  \hline 
  Blaze TV  &  0.02  & 0.05 & -0.03 \\
    \hline
  
    \end{tabular}
\vspace{0.2cm}  
\caption{Analysis of viewership agreement. For a given news network and time duration, each cell summarizes the $\frac{\Sigma_i \mathbbm{I}(v^i,  \mathcal{C}, \mathcal{T}) \frac{v^i_\mathit{dislike}}{v^i_\mathit{dislike} + v^i_\mathit{like}}}{\Sigma_i \mathbbm{I}(v^i, \mathcal{C}, \mathcal{T})}$. $\mathcal{T}~\in~\{\mathcal{T}_\mathit{before}, \mathcal{T}_\mathit{after}\}$; $\mathbbm{I}(v^i, \mathcal{C}, \mathcal{T})$ returns 1 if video $v^i$ is uploaded within duration $\mathcal{T}$, otherwise it returns 0.}
\label{tab:agreement}
\end{center}
}
\end{table}

\subsubsection{Comments on videos}

The two time slices we are focusing on, both are expected to generate high news viewership in our current political climate. $\mathcal{T}_{\mathit{before}}$, i.e., the time slice leading up to the election would naturally attract viewers because of the coverage of political debates, rally speeches, and election predictions. As a result of casting widespread doubts over the legitimacy of the election, we anticipated the engagement during $\mathcal{T}_{\mathit{after}}$ would be high as well. Also, note that, since any video uploaded during $\mathcal{T}_{\mathit{before}}$ would have more time to accrue comments than any video uploaded during $\mathcal{T}_{\mathit{after}}$, it is not surprising if the average number of comments for videos uploaded during $\mathcal{T}_{\mathit{after}}$ is slightly less than the average number of comments for videos uploaded during $\mathcal{T}_{\mathit{after}}$ for a given channel. However, Table~\ref{tab:engagement} shows three distinct patterns. We notice that (1) CNN, Blaze TV and MSNBC do not show any noticeable change in average number of comments; (2) Fox News shows noticeable decline in comment engagement; and (3) OANN and Newsmax show an increase by more than factors of 2 and 3, respectively.   

\begin{table}[htb]
{

\begin{center}

\begin{tabular}{|l | c | c |}
    \hline
    YouTube channel & $\mathcal{T}_\mathit{before}$ & $\mathcal{T}_\mathit{after}$\\
    \hline                                 
   CNN   &  398~/~4,188	      &  426~/~3,993 \\
    \hline
   Fox News &  1,040~/~2,587  & 1,026~/~1,809\\
    \hline
  MSNBC  &  1,891~/~709  & 1,999~/~719 \\
    \hline
  OANN  &  1,090~/~163    & 638~/~\textbf{399} \\
  \hline
  Newsmax  &  294~/~453 & 452~/~\textbf{1,852} \\
  \hline 
  Blaze TV  &  280~/~1,241  & 238~/~1,206 \\
    \hline
  
    \end{tabular}
\end{center}
}
\caption{Analysis of engagement. For a given news network and time duration, each cell summarizes the channel activity as \textbf{$a$} / \textbf{$b$} where \textbf{$a$} denotes the number of videos uploaded and \textbf{$b$} denotes the average number comments on videos where commenting is allowed. We note that Newsmax and OANN enjoyed a remarkable increase in average number of comments per video. Note that, OANN was banned for a week by YouTube because of spreading COVID-19 misinformation.}
\label{tab:engagement}
\end{table}

\subsubsection{Number of subscribers} Comments on a news video or likes or dislikes are response to an individual unit of content supplied by a given channel -- a single video. YouTube viewers can subscribe to specific channels indicating that they are interested in receiving updates on the channel's activities (e.g., receive notification when a new video is uploaded). In that sense, subscription to a channel is perhaps a more longer term engagement signal than liking (or disliking) or commenting. Let for each channel $\mathcal{C}$, $\mathcal{C}^{t, \mathit{sub}}$ denote the total number of subscribers of $\mathcal{C}$ at time $t$. We define market-share of subscribers of a given channel $\mathcal{C}_i$ at time $t$ as:\\
$\mathit{marketShare}(\mathcal{C}_i,t) = \frac{\mathcal{C}_{i}^{t, \mathit{sub}}}{\Sigma_j \mathcal{C}_{j}^{t, \mathit{sub}}}$ where $\mathcal{C}_i, \mathcal{C}_j \in \{\mathit{Newsmax}, \mathit{Blaze}, \mathit{CNN}, \mathit{OANN}, \mathit{Fox}, \mathit{MSNBC}\} $. We admit that our definition oversimplifies certain things since a specific user can subscribe to multiple news networks at the same time.  Also, there can be several other possible news sources even on YouTube. That said, our measure allows us to track the growth of these six networks revealing insights into the nature of  growth of these fringe networks in the last 128 days. 

\begin{figure}[t]
\centering
\includegraphics[trim={0 0 0 0},clip, height=1.8in]{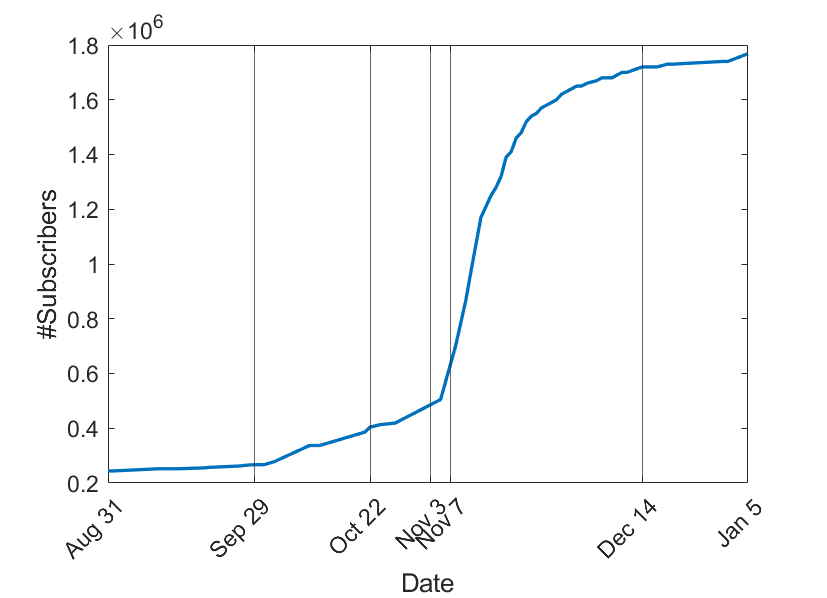}
\caption{{The growth of Newsmax in terms of \#subscribers. The vertical lines indicate important dates. The two Presidential debates took place on September 29$^\mathit{th}$} and October 22$^\mathit{nd}$.
The election took place on 3$^{rd}$ November and AP called the election for Biden on 7$^\mathit{th}$ November. The electoral college vote took place on December 14$^\mathit{th}$.}
\label{fig:newEngagement}
\end{figure}

Table~\ref{tab:marketshare} summarizes the market-share of each of the news networks on three particular days: (1) 31$^\mathit{st}$ August 2020, the first day of $\mathcal{T}_\mathit{before}$; (2) 3$^\mathit{rd}$ November, 2020, the first day of $\mathcal{T}_\mathit{after}$ and the day of 2020 US election; and (3) January 5$\mathit{th}$, 2021, the last day of $\mathcal{T}_\mathit{after}$\footnote{We obtained the \#subscribers from \url{https://web.archive.org/}. Piecewise linearity is assumed for missing entries.}. We note that (1) all fringe news networks gained market-share as time progressed with Newsmax's gain being equal to a factor of 6 (Figure~\ref{fig:newEngagement} presents its growth in subscriber-count); (2) the big-three (CNN, Fox News, and MSNBC) lost market share when we compare their individual market-shares on 5$^\mathit{th}$ January, 2021 with what was on 31$^\mathit{st}$ August, 2020; and (3) Fox News exhibits a curious pattern where the market-share slightly rises on 3$^{rd}$ November, 2020 and then dips.    

\begin{table}[t]
{

\begin{center}

\begin{tabular}{|l | c | c | c |}
    \hline
    YouTube channel & 31$^\mathit{st}$ August, 2020 & 3$^\mathit{rd}$ November, 2020 & 5$^\mathit{th}$ January, 2021 \\
    \hline                                 
   CNN   &  47.63\%	     &46.14\% &  43.67\% \\
    \hline

   Fox News &  27.44\%  & 27.64\% & 24.97\%\\
    \hline

  MSNBC  &  15.61\% &15.24\% & 14.72\% \\
    \hline

  OANN  &   3.75\%    &3.97\% & \textbf{5.04\%} \\
  \hline
   
  Newsmax   & 1.11\% &2.02\% & \textbf{6.60\%} \\
  \hline 
  Blaze TV  &  4.46\%  &4.99\% & \textbf{5.00\%} \\
    \hline
  
    \end{tabular}
\vspace{0.2cm} 
\caption{Analysis of market-share in terms of subscriber count. We define the market-share of a channel in a particular time $t$ as the ratio of  its subscriber count to the sum of subscriber count at time $t$ of all the news channels considered.}
\label{tab:marketshare}
\end{center}
}
\end{table}

\subsection{User Migration}\label{sec:fox}

\noindent\emph{\textbf{RQ 4}: Was there any systematic migrations from mainstream media outlets to fringe media outlets?}

\begin{table}[htb]
{

\begin{center}

\begin{tabular}{|l | c | c | c | c |}
    \hline
    YouTube channel & $\mathcal{T}_\mathit{before}$ & $\mathcal{T}_\mathit{after}$ & $\mathcal{T}_\mathit{earliest}$ & $\mathcal{T}_\mathit{latest}$\\
    \hline                                 
   Fox News - Newsmax   &  91\%~/~9\%	      &  57\%~/~43\% &89\%~/~11\%	  &59\%~/~41\%	  \\
    \hline
   CNN - MSNBC &  45\%~/~55\%  & 46\%~/~54\%&45\%~/~55\%&47\%~/~53\%\\
   
       \hline
  
    \end{tabular}
\vspace{0.2cm} 
\caption{Analysis of comment-share between pair of networks. For a network pair $\langle\mathcal{C}_1, \mathcal{C}_2\rangle$ the share is summarized as \textbf{$a$} / \textbf{$b$} where \textbf{$a$} denotes comment share of $\mathcal{C}_1$  and \textbf{$b$} denotes comment share of $\mathcal{C}_2$.}
\label{tab:migration}
\end{center}
}
\end{table}

Table~\ref{tab:engagement}, Table~\ref{tab:agreement}, and Table~\ref{tab:marketshare} all point to a decline in Fox News's popularity during $\mathcal{T}_\mathit{after}$ as compared to $\mathcal{T}_\mathit{before}$. We are curious to examine where did these viewers go? 
Let $\mathcal{N}_\mathit{fox}^i$ and $\mathcal{N}_\mathit{newsmax}^i$ denote the total number of comments made by user $u_i$ on Fox News videos and Newsmax videos uploaded during $\mathcal{T}_\mathit{128}$, respectively. 
We focus on highly active users who commented both on Fox News and Newsmax videos to obtain a user set $\mathcal{U}$ such that $u_i \in \mathcal{U}$ iff $\mathcal{N}_\mathit{fox}^i > 0$, $\mathcal{N}_\mathit{newsmax}^i > 0$ and $\mathcal{N}_\mathit{fox}^i + \mathcal{N}_\mathit{newsmax}^i \ge 10$. In plain words, our user set contains users who have made at least one comment on Fox News and Newsmax and the total number of comments made on Fox News and Newsmax by the user exceeds or equals 10.  We obtain 69,766 users satisfying these conditions. We then analyze their activities by slicing $\mathcal{T}_\mathit{128}$ in two different ways. One natural choice is the temporal slices $\mathcal{T}_\mathit{after}$ and $\mathcal{T}_\mathit{before}$. Our second choice of time slice divided $\mathcal{T}_\mathit{128}$ along the activity timeline of a given user. We consider the earliest 20\% and the latest 20\% comments made by each user during $\mathcal{T}_\mathit{128}$ and analyze the relative share of comments in Fox News and Newsmax. 

Table~\ref{tab:migration} summarizes our findings. In order to contrast our results, as a control group, we consider the channel pair of CNN and MSNBC and analyzed the comment shared of user group of 99,101 users following the conditions described above. We notice that during the distribution of comments in CNN-MSNBC pair was stable across $\mathcal{T}_\mathit{before}$ and $\mathcal{T}_\mathit{after}$. However, we notice a stark contrast in Fox-Newsmax pair. During $\mathcal{T}_\mathit{before}$, Newsmax has a minuscule presence while $\mathcal{T}_\mathit{after}$ exhibits a near equal comments share with Fox. The qualitative trend of this analysis remains unchanged even when we consider our user activity-based timeline. Hence, our analyses indicate that indeed, several users from Fox News moved to Newsmax.

\subsection{Cloze Tests}\label{sec:cloze}

\noindent\emph{\textbf{RQ 2}: Did the audience of different networks exhibit different attitudes toward accepting and presenting the election outcomes?}

We now investigate this research question through the lens of cloze tests using language models. The masked word prediction of high-performance language models, such as \texttt{BERT}~\cite{devlin2018bert}, has a parallel in the form of cloze tests~\cite{taylor1953cloze} aka fill-in-the-blank questions used in the human psycholinguistics literature~\cite{smith2011cloze}. When presented with a sentence (or a sentence stem) with a missing word, a cloze task is essentially a fill-in-the-blank task. For instance, in the following cloze task: \emph{In the} \texttt{[MASK]}\emph{, it snows a lot}, \texttt{winter} is a likely completion for the missing word. In fact, when given this cloze task to \texttt{BERT}, \texttt{BERT} outputs the following five seasons ranked by decreasing probability: \texttt{winter}, \texttt{summer}, \texttt{fall}, \texttt{spring} and \texttt{autumn}. In a different political context of the 2019 Indian general election,~\cite{electionKhudaBukhsh} has demonstrated that \texttt{BERT} can be fine-tuned on large-scale social media political discussions to efficiently aggregate political opinions and track evolving national priorities through simple cloze tests like   \emph{The biggest problem of India is} \texttt{[MASK]}. 

In our work, we are interested in gauging the aggregate attitude of a network viewership toward the outcome of the 2020 election. For each channel, we fine-tune \texttt{BERT} with the comments on videos uploaded during $\mathcal{T}_\mathit{after}$. \texttt{BERT}'s vulnerability in handling negations is documented in~\cite{kassner-schutze-2020-negated}.  Following~\cite{electionKhudaBukhsh}, we remove all comments that contains any valence shifter. 

Before presenting our results on the aggregate opinion of each of the networks' viewership on the election, we make a small digression to discuss a result that sheds light on the stark contrast of opinions across these news networks. On a cloze test \emph{The biggest problem of America is} \texttt{[MASK]}, we notice that the top three results  succinctly capture the divergent views of the news audience across news networks. While $\mathit{socialism}$ consistently appeared in all conservative networks, $\mathit{trump}$, $\mathit{covid}$, and $\mathit{racism}$ appeared in the their liberal counterparts.      

To rank the aggregate opinion on the 2020 US election, we consider the following two cloze tests: (1) \emph{Trump has} \texttt{[MASK]}~\emph{the 2020 election} (denoted by $\mathit{cloze}_\mathit{trump}$) (2) \emph{Biden has}  \texttt{[MASK]}~\emph{the 2020 election} (denoted by $\mathit{cloze}_\mathit{biden}$). Let $\mathit{clozeTest}(\mathit{c}, w)$ denote the probability of the word $w$ output by \texttt{BERT}. In order to appropriately calibrate the model, we compute the score for Trump as 
\large{$\frac{\mathit{clozeTest}(\mathit{cloze}_\mathit{trump}, won)}{\mathit{clozeTest}(\mathit{cloze}_\mathit{trump}, won) + \mathit{clozeTest}(\mathit{cloze}_\mathit{biden}, won)}$}\\
\normalsize 
and Biden as \large{$\frac{\mathit{clozeTest}(\mathit{cloze}_\mathit{biden}, won)}{\mathit{clozeTest}(\mathit{cloze}_\mathit{trump}, won) + \mathit{clozeTest}(\mathit{cloze}_\mathit{biden}, won)}$}.
\normalsize
Note that, for any channel, the scores for Trump and Biden sum to 1. The scores for Trump for different news networks give us the following order: $\mathit{MSNBC}$ < $\mathit{CNN}$ < $\mathit{Fox}$ < $\mathit{OANN}$ < $\mathit{Blaze}$ < $\mathit{Newsmax}$. This result indicates that compared to mainstream media outlets, discussions on fringe news channels exhibit more doubts on the legitimacy of the election.   

\begin{table*}[htb]
\small
{
\begin{center}
     \begin{tabular}{ |p{2cm} | p{2cm} | p{2cm} | p{2cm} | p{2cm} |
     p{2cm} |}
    \hline
  MSNBC    
                                                 & CNN                                               & Fox News                                                      & OANN                           & Newsmax & Blaze TV                                            \\ \hline
         
$\mathit{trump}$, $\mathit{covid}$, $\mathit{umemployment}$ & $\mathit{corruption}$, $\mathit{racism}$, $\mathit{trump}$ & 
$\mathit{socialism}$, $\mathit{corruption}$, $\mathit{communism}$
& $\mathit{communism}$, $\mathit{corruption}$, $\mathit{socialism}$ & $\mathit{communism}$, $\mathit{corruption}$, $\mathit{socialism}$ & $\mathit{corruption}$, $\mathit{poverty}$, $\mathit{socialism}$ \\ \hline

    \end{tabular}
\end{center}

\label{tab:allCloze}}
\caption{{Cloze test results for the probe \emph{The biggest problem of America is} \texttt{[MASK]}. A separate version of \texttt{BERT} was fine-tuned, using viewer comments from each network. Top three results (ranked by probability) output by fine-tuned \texttt{BERT} are presented for each news network.}}
\end{table*}

We further corroborated our results with a well-known natural language inference model~\cite{Gardner2017AllenNLP}. Given a premise text and a hypothesis text, the natural language inference (NLI) task is to predict either entailment, contradiction, or  independence. For example, the hypothesis \emph{some men are playing a sport} is entailed by the premise \emph{a soccer game with multiple males playing}~\footnote{This example is taken from~\cite{bowman-etal-2015-large}}. Our work draws inspiration from a recent work~\cite{hossain-etal-2020-covidlies} that cast the task of COVID-19 misinformation detection as an NLI task stating that the class labels \emph{informative}, \emph{misinformative} and \emph{irrelevant} has a natural one-to-one correspondence to \emph{entailment}, \emph{contradiction} and \emph{semantic irrelevance},  respectively. For a given news network, using individual comments from our data set as premise, we considered the following two hypotheses:

\noindent\emph{\textbf{Hypothesis 1}: I prefer Trump as my president.} (denoted by $\mathcal{H}_1$)\\
\noindent\emph{\textbf{Hypothesis 2}: I prefer Biden as my president.} (denoted by $\mathcal{H}_2$)

For a given channel $\mathcal{C}$ and a hypothesis $\mathcal{H}$, we randomly sampled 5,000 comments from user discussions on videos uploaded by $\mathcal{C}$ during $\mathcal{T}_\mathit{after}$ and compute the fraction of comments that entail $\mathcal{H}$ using an off-the-shelf, well-known NLI inference system~\cite{Gardner2017AllenNLP}.

\noindent Obtained order for $\mathcal{H}_1$, from least to greatest: 
$\mathit{MSNBC}$ <
$\mathit{CNN}$ < $\mathit{Fox}$ < $\mathit{OANN}$ < $\mathit{Newsmax}$ < $\mathit{Blaze}$.

\noindent Obtained order for $\mathcal{H}_2$ from least to greatest: 
$\mathit{Blaze}$ < $\mathit{OANN}$ < $\mathit{Fox}$ < $\mathit{Newsmax}$ < $\mathit{MSNBC}$ < $\mathit{CNN}$.

\subsection{Machine Translation Based Analysis}\label{sec:translation}

\noindent\emph{\textbf{RQ 5}: Based on the comments on the viewed videos, which news networks were ``linguistically most similar'' to those of President Trump's YouTube channel?}

Quantifying the differences between large-scale social media discussion data sets is a challenging task and we recourse to the most-recent method in the literature~\cite{khudabukhsh2020dont}. In~\cite{khudabukhsh2020dont}, the authors presented a machine translation based framework. This framework assumes that two sub-communities (e.g., Fox viewers and CNN viewers) are speaking in two different \emph{languages} (say, $\mathcal{L}_\mathit{cnn}$ and $\mathcal{L}_\mathit{fox}$) and obtains single-word translations using a well-known machine translation algorithm~\cite{SmithTHH17}. In a world not fraught with polarization, any word $w$ in $\mathcal{L}_\mathit{cnn}$ should translate to itself in $\mathcal{L}_\mathit{fox}$. However, if a word $w_1$ in one language translates to a different word $w_2$ in another, it indicates $w_1$ and $w_2$ are used in similar contexts across these two \emph{languages} signalling (possible) disagreement. These disagreed pairs\footnote{The original paper~\cite{khudabukhsh2020dont} refers to these pairs as misaligned pairs.} present a quantifiable measure to compute differences between large scale corpora as greater the number of disagreed pairs the farther two sub-communities are. 

Formally, let our goal be computing the similarity measure between two languages, $\mathcal{L}_\mathit{source}$ and $\mathcal{L}_\mathit{target}$, with vocabularies $\mathcal{V}_\mathit{source}$ and $\mathcal{V}_\mathit{target}$, respectively. Let $\emph{translate}(w)^{\mathcal{L}_\mathit{source} \rightarrow \mathcal{L}_\mathit{target}}$ denote a single word translation of $w \in \mathcal{V}_\mathit{source}$ from $\mathcal{L}_\mathit{source}$ to $\mathcal{L}_\mathit{target}$. The similarity measure between two languages along a given translation direction computes the fraction of words in $\mathcal{V}_\mathit{source}$  that translates to itself, i.e.,\\
\large{
\emph{Similarity}\hspace{0.03cm}($\mathcal{L_\mathit{source}}, \mathcal{L_\mathit{target}})$ = $\frac{\Sigma_{w \in \mathcal{V}_\mathit{source}} \mathbbm{I}(\emph{translate}(w)^{\mathcal{L}_\mathit{source} \rightarrow \mathcal{L}_\mathit{target}} = w)}{|\mathcal{V}_\mathit{source}|}$
}.
\normalsize
The indicator function returns 1 if the word translates to itself and 0 otherwise. The larger the value of \emph{Similarity} ($\mathcal{L_\mathit{source}}, \mathcal{L_\mathit{target}})$, the greater is the similarity between a language pair.  

Beyond prominent US cable news networks,~\cite{khudabukhsh2020dont} has computed similarities between news networks and discussions on YouTube videos hosted by major prime time US political comedians. In this work, we turn our focus to President Trump whose official YouTube handle has 2.68 million subscribers as of 5$^\mathit{th}$ January, 2021 (see, Table~\ref{tab:network}). We follow the same steps and hyper-parameter settings described in~\cite{khudabukhsh2020dont} and in Table~\ref{tab:translation}, we quantify the similarities between language present in the official YouTube channel of the 45$^\mathit{th}$ US president  (denoted by $\mathcal{L}_\mathit{trump}$) and the six US cable news networks. We use the same monikers for the languages in the four news networks considered in~\cite{khudabukhsh2020dont} ($\mathcal{L}_\mathit{cnn}$, $\mathcal{L}_\mathit{fox}$, $\mathcal{L}_\mathit{msnbc}$, and $\mathcal{L}_\mathit{oann}$) and denote the language of the discussions on Blaze TV and Newsmax news videos as $\mathcal{L}_\mathit{blaze}$ and $\mathcal{L}_\mathit{newsmax}$, respectively.

\begin{table}[t]
\centering
\small
\setlength{\extrarowheight}{2pt}
\begin{tabular}{cc|c|c|c|c|c|c|c|c|c|c|c|c|c|}
  & \multicolumn{1}{c}{} & \multicolumn{7}{c}{$\mathcal{L}_\mathit{target}$} \\
  & \multicolumn{1}{c}{} &
  \multicolumn{1}{c}{$\mathcal{L}_\mathit{trump}$}
  &\multicolumn{1}{c}{$\mathcal{L}_\mathit{cnn}$}  & \multicolumn{1}{c}{$\mathcal{L}_\mathit{fox}$}  & \multicolumn{1}{c}{$\mathcal{L}_\mathit{msnbc}$} & 
  \multicolumn{1}{c}{$\mathcal{L}_\mathit{oann}$} & 
  \multicolumn{1}{c}{$\mathcal{L}_\mathit{newsmax}$} &
  \multicolumn{1}{c}{$\mathcal{L}_\mathit{blaze}$} 
  \\\cline{3-9}
          \multirow{10}{*}    & $\mathcal{L}_\mathit{trump}$ & \cellcolor{blue!25} - & \cellcolor{gray!25}43.8\%  & \cellcolor{gray!25}47.0\%  & \cellcolor{gray!25}39.5\%  & 
      \cellcolor{gray!25}48.6\%
          & \cellcolor{gray!25}\textbf{49.2\%} & \cellcolor{gray!25}47.3\% \\ \cline{3-9}
            & $\mathcal{L}_\mathit{cnn}$ &42.5\%  & \cellcolor{blue!25}- &72.4\%   &77.2\%  &56.3\%&46.4\%&60.1\% \\ \cline{3-9}
            & $\mathcal{L}_\mathit{fox}$ &46.4\%  &71.4\%  & \cellcolor{blue!25}-   &66.7\% &67.6\% &57.3\% &74.2\%\\ \cline{3-9}
        {$\mathcal{L}_\mathit{source}$}    & $\mathcal{L}_\mathit{msnbc}$ &39.6\%  &79.4\%  &67.1\%  & \cellcolor{blue!25}- &53.3\%&47.8\%&54.5\% \\ \cline{3-9}
            &  $\mathcal{L}_\mathit{oann}$ & 46.9\%   &56.0\%   &68.5\%  &53.4\%  &\cellcolor{blue!25} -&62.4\%& 73.9\% \\\cline{3-9} 
             & $\mathcal{L}_\mathit{newsmax}$ & 48.3\%  &46.9\%    & 58.5\%  & 47.5\%   & 62.4\%& \cellcolor{blue!25}-&62.1\% \\\cline{3-9} 
              & $\mathcal{L}_\mathit{blaze}$ &46.8\%&60.1\%  &74.1\%  &54.6\%  &72.6\% &61.7\%&  \cellcolor{blue!25} - \\\cline{3-9} 
     
\end{tabular}
\vspace{0.4cm}
\caption{{Pairwise similarity between news-languages and $\mathcal{L}_\mathit{trump}$ computed for the year 2020 using the framework presented in~\cite{khudabukhsh2020dont}. The cells show the similarity between the language pair along the translation direction of language shown in the row as source and the language shown in the column as target. Hyper-parameters are identical to~\cite{khudabukhsh2020dont}.  $\mathcal{L}_\mathit{trump}$ (relevant row-cells  are shaded with gray) is found to be most similar with $\mathcal{L}_\mathit{newsmax}$. Appendix contains additional experiments focusing on $\mathcal{T}_\mathit{after}$ and considering $\mathcal{L}_\mathit{fox}$, $\mathcal{L}_\mathit{newsmax}$ and $\mathcal{L}_\mathit{trump}$.}}
\label{tab:translation}

\end{table}

It is well-known that corpus size is one of the most important contributing factors to ensure the quality of word embedding~\cite{NIPS2013_5021}. Further,~\cite{khudabukhsh2020dont} indicates that typical to most deep learning systems, one of the limitations of the machine translation based framework is it is data-hungry. We thus focus on the entire year of 2020 (data set details are provided in the Appendix). Table~\ref{tab:translation} underscores the following two points: (1) the language present in the YouTube videos hosted by the official channel of President Trump is \textbf{more similar} to fringe media outlets than any mainstream media outlet with the ordering (most similar to least similar): $\mathcal{L}_\mathit{newsmax}$ > $\mathcal{L}_\mathit{oann}$ > $\mathcal{L}_\mathit{blaze}$ > $\mathcal{L}_\mathit{fox}$ > $\mathcal{L}_\mathit{cnn}$ > $\mathcal{L}_\mathit{msnbc}$; and and (2) compared to the liberal news outlets, the conservative news networks are more similar to each other. Also, note that, the 45$^\mathit{th}$ US president's YouTube channel is not a news network. Hence, it does not cover issues as varied as a typical news network would. Therefore, it is not surprising that the similarity between $\mathcal{L}_\mathit{trump}$ and other news-languages are lesser than the similarity between news networks.

\section{Discussions and Conclusions}

\subsection{Discussions}
\noindent \textbf{{A mysterious 11-character word:}} On a Skip-gram word embedding~\cite{bojanowski2017enriching} trained on discussions from $\mathcal{T}_\mathit{after}$ for a specific channel, we noticed a curious 11-character word among the nearest neighbors of the phrase \emph{voter fraud}. Upon examination, we realized that it is a YouTube video ID. Soon we realized that when restricted to a specific character length of 11, nearest neighbors of \emph{voter fraud} in the word embedding space reveal several video IDs, most of which cast doubts on the fairness of the election. Not only that, we found that nearest neighbors of a video ID of a video propagating voter fraud misinformation are also video IDs of videos with similar content. In this intriguing  phenomenon where distributional hypothesis~\cite{harris1954distributional} meets misinformation, we were surprised to notice the wide range of viewer-reach these videos possessed. Of the 30 nearest neighbors we manually annotated, 28 cast doubts about the electoral process and the viewer-count ranged from a paltry 105 to more than a million views. Our findings  indicate that during this political crisis, it is possible that beyond these high-traffic news networks and influencers, several other videos promoting unsubstantiated claims surfaced in the comments section of a mainstream social media platform and it is a challenging task to catch them all. 

\noindent \textbf{{Consumption pattern:}} While in this work we focus on the fringe media outlets, instead of conservative forums such as parler or gab, our choice of the platform could not be more mainstream: YouTube. Beyond YouTube's tremendous popularity in the US (126 million unique US users in 2020 according to Statista) YouTube is compelling platform for another reason. Because YouTube offers access to these different networks through a single, uniform interface, it is easy for consumers to effectively “flip channels”, and easy to track individual behavior from YouTube data, making it an ideal platform for our study. Detecting anomalous consumption patterns such as the abrupt rise of Newsmax in popularity  is a much easier task than automatically identifying presence of unsubstantiated claims from videos. Our work thus raises an important point that during a political crisis, consumption patterns may reveal useful signals.  


%

\noindent \textbf{{Internet abhors vacuum:}} Our work is an important study in the context of this unique crisis to Western democracy which shows that with the current almost-ubiquitous penetration of the internet, vacuums may fill up rapidly. If a mainstream media is unwilling to present an \emph{alternate  version} of the election outcome, certain fringe networks can fill up the void and enjoy a sudden meteoric rise in popularity possibly through presenting an \emph{alternate version} of reality. As compared to OANN and Newsmax, the rise in popularity of Blaze TV was relatively muted. While the content and audience of this network is not much different from the other two fringe networks, the 45$^\mathit{th}$ President of the US  tweeted favorably about OANN and Newsmax on multiple occasions. Our analysis cannot present causal evidences. Neither can it rule out the possibility that a different fringe network will not enjoy a similar run as Newsmax in a subsequent political crisis in the near future.   

\subsection{Conclusions}

This paper leads to two different types of conclusions: conclusions about what actually transpired during the 128 days covered by our data set, and conclusions about methodologies for analyzing such large scale social media to study political and other social sciences.

\subsubsection{Understanding What Happened}

Through a series of corroborating experiments described above, we make the following conclusions.  

\noindent\emph{\textbf{C1}: Fringe networks did not cover the election the same way as the mainstream networks.}
(\textbf{\emph{RQ 1}}, Section~\ref{sec:trigram}). 

\noindent\emph{\textbf{C2}: Audience of the fringe networks exhibit more doubt about the election outcome as compared to the audience of mainstream outlets.}
(\textbf{\emph{RQ 2}}, Section~\ref{sec:trigram}; \textbf{\emph{RQ 2}}, Section~\ref{sec:cloze})

\noindent\emph{\textbf{C3}: A subset of fringe news networks gained audience post election.} (\textbf{\emph{RQ 3}}, Section~\ref{sec:engagement}; \textbf{\emph{RQ 4}}, Section~\ref{sec:fox})

\noindent\emph{\textbf{C4}: Fox News was the only mainstream media outlet that lost considerable popularity.} (\textbf{\emph{RQ 3}}, Section~\ref{sec:engagement}; \textbf{\emph{RQ 4}}, Section~\ref{sec:fox})

\noindent\emph{\textbf{C5}: Viewer comments on President Trump's official YouTube handle are linguistically more similar to viewer comments on fringe networks than those of the more mainstream media outlets.} (\textbf{\emph{RQ 5}}, Section~\ref{sec:translation})


\subsubsection{Methodology} We demonstrate that recent advancements in NLP methods enable us to analyze a vast amount of data in almost real time with minimal manual supervision. However, each of these methods has certain blindspots (e.g., \texttt{BERT}'s vulnerability to negation or the translation based method's requirement of a large of amount data). Our work demonstrates the synergy of these methods in obtaining corroborating evidences from multiple sources and thus gaining valuable insights. While these techniques have been used in isolation on different political corpora~\cite{electionKhudaBukhsh, khudabukhsh2020dont,hossain-etal-2020-covidlies}, in this work, we present a combined approach to analyze a data set on a political crisis the country has not seen for years.

\bibliographystyle{unsrt}

\newpage
\section{Appendix}

\subsection{Experimental Setup}
Experiments are conducted in a suite of machines with the following specifications: 
\begin{compactitem}
\item OS: Windows 10.
\item Processor	Intel(R) Core(TM) i7-9750H CPU @ 2.60GHz, 2592 Mhz, 6 Core(s), 12 Logical Processor(s).
\item RAM: 64 GB.
\end{compactitem}

\subsection{Preprocessing and Hyperparameters}

To train word embedding on our data set, we use the following preprocessing steps. First, we remove all the emojis and non-ascii characters. Then, we remove all non-alphanumeric characters and lowercase the remaining text. We preserve the newline character after each individual document in the data set. We use the default parameters for training our FastText~\cite{bojanowski2017enriching} Skip-gram embedding with the dimension set to 100.

\subsection{Machine Translation Based Analysis}

\begin{table}[htb]
\centering
\scriptsize
\setlength{\extrarowheight}{2pt}
\begin{tabular}{cc|c|c|c|}
  & \multicolumn{1}{c}{} & \multicolumn{3}{c}{$\mathcal{L}_{\emph{target}}$} \\
  & \multicolumn{1}{c}{} & \multicolumn{1}{c}{$\mathcal{L}_{\emph{trump}}$}  & \multicolumn{1}{c}{$\mathcal{L}_{\emph{fox}}$}  & \multicolumn{1}{c}{$\mathcal{L}_{\emph{newsmax}}$} \\\cline{3-5}
            & $\mathcal{L}_{\emph{trump}}$ &\cellcolor{blue!25} - & 35.6\%   & 49.0\% 
 \\ \cline{3-5}
$\mathcal{L}_{\emph{source}}$  & $\mathcal{L}_{\emph{fox}}$ & 36.0\%  &\cellcolor{blue!25} - &  58.6\%
 \\\cline{3-5}
            & $\mathcal{L}_{\emph{newsmax}}$ &47.5\%  &57.8\%   &\cellcolor{blue!25} - \\\cline{3-5}
\end{tabular}
\vspace{0.5cm}
\caption{{Pairwise similarity between languages computed for videos uploaded during $\mathcal{T}_\mathit{after}$. Hyperparameters are identical to~\cite{khudabukhsh2020dont}. The cells show the similarity between the language pair along the translation direction of language shown in the row as source and the language shown in the column as target.}}
\label{tab:threeChannels}
\end{table}

Table~\ref{tab:threeChannels} indicates that if we zoom onto videos uploaded during $\mathcal{T}_\mathit{after}$, the qualitative claim that linguistically $\mathcal{L}_\mathit{trump}$ is more similar to $\mathcal{L}_\mathit{fox}$ than $\mathcal{L}_\mathit{newsmax}$, still holds. 

\subsection{Data Set Details for 2020}

Table~\ref{tab:2020} summarizes the details of our extended data set. 

\begin{table}[htb]
{
\caption{Data set details for 2020.}
\begin{center}
     
\begin{tabular}{|l | c  |c |}
    \hline
    YouTube Channel & \#Videos  &\# Overall comments  \\
    \hline                                 
   CNN    &  2,973  &9.27M \\
    \hline
   Fox News  & 6,066  &11.6M  \\
    \hline
  MSNBC  &  10,644  &6.08M  \\
    \hline
  OANN   &  5,092  & 1.01M \\
  \hline
  Newsmax   &  1,673  & 1.05M \\
  \hline 
  Blaze TV   &  1,353  & 1.36M \\
  \hline
  Donald J. Trump   & 3,991  & 2.7M \\
  \hline
    \end{tabular}

\label{tab:2020}
\end{center}
}
\end{table}

\end{document}